# Vacancy formation energies and migration barriers in multi-principal element alloys

Ankit Roy[a,b],* Prashant Singh[b],* Ganesh Balasubramanian[a], and Duane D. Johnson[b, c]

[a] Department of Mechanical Engineering and Mechanics, Lehigh University, Bethlehem, PA 18015, USA
[b] Ames Laboratory, United States Department of Energy, Iowa State University, Ames, IA 50011, USA
[c] Department of Materials Science & Engineering, Iowa State University, Ames, IA 50011, USA

**Abstract**

Multi-principal element alloys (MPEAs) continue to garner interest as structural and plasma-facing materials due to their structure stability and increased resistance to radiation damage. Despite sensitivity of mechanical behavior to irradiation and point-defect formation, there has been scant attention on understanding vacancy stability and diffusion in refractory-based MPEAs. Using density-functional theory, we examine vacancy stability and diffusion barriers in body-centered cubic $(Mo_{0.95}W_{0.05})_{0.85}Ta_{0.10}(TiZr)_{0.05}$. The results in this MPEA show strong dependence on environment, originating from local lattice distortion associated with charge-transfer between neighboring atoms that vary with different chemical environments. We find a correlation between degree of lattice distortion and migration barrier: (Ti, Zr) with less distortion have lower barriers, while (Mo, W) with larger distortion have higher barriers, depending up local environments. Under irradiation, our findings suggest that (Ti, Zr) are significantly more likely to diffuse than (Mo, W) while Ta shows intermediate effect. As such, material degradation caused by vacancy diffusion can be controlled by tuning composition of alloying elements to enhance creep strength at extreme operating temperatures and harsh conditions.

# 1. Introduction

Power-generation plants, both thermal and nuclear, now demand novel structural and coating materials that retain excellent mechanical properties and stability even at elevated temperatures (>1000°C). This quest for higher operating temperatures has led to a surge in research on complex alloys to achieve higher efficiency in power-generation technologies [1]. With this as motivation, the concept of multi-principal element alloys (MPEAs) was introduced by Yeh [2] and Cantor [3], with high-entropy alloys limited to near-equiatomic compositions. MPEAs have created a global revolution in alloy design never witnessed in past metallurgical research [2-5]. MPEAs were originally defined as having five or more elements in relatively high concentrations (5-35%). To enhance mechanical properties like hardness and elastic modulus at elevated temperatures [2, 6-8], the principle was to add more elements and to maximize the configurational entropy to favor formation of simpler single-phase alloys [2, 9]. But, in the past several years, research on metastable MPEAs has revealed many alloys violate this principle [10-12]. Hence, the goal shifted towards developing alloys that may not contain as many elements in equal proportions, but rather a complex mix in an optimized fashion to obtain the best properties [4, 13], including in medium-entropy (3 or 4 elements) alloys. In the quest for promising compositions, experiments alone would do very little to explore the enormous search space with several billion MPEAs [14]. More recently, density-functional theory (DFT) [10,14] or molecular dynamics (MD) [15-17] and combined with machine-learning [7, 8, 18, 19] techniques played a key role in accelerating the discovery of promising MPEAs.

Currently, tungsten (W) is the leading candidate for plasma-facing components (PFCs) owing to its high melting temperature of 3422°C, high thermal conductivity, low tritium retention, and low sputtering-erosion rates in cold scrape-off layer plasma [20, 21]. Despite its widespread application in PFCs, tungsten has limitations restricting its use in fusion reactors, like poor fracture toughness, high brittle-to-ductile transition temperature, and a risk of forming nanoscale bubbles under irradiation by He ions [22]. As such, one alternative route is by tuning MPEA properties for applications where good irradiation resistance is required. In fact, recent experimental efforts have revealed certain body-centered cubic (bcc) alloys like $W_{38}Ta_{36}Cr_{15}V_{11}$ show no trace of dislocation loops created during irradiation up to a dose of 8 displacements-per-atom (dpa) [23]. Unlike as observed in pure W, the mobility of self-interstitials and vacancies is similar and there is a large imbalance in defect mobility [24, 25]. Apart from excellent irradiation resistance, refractory-based

MPEAs with bcc phase exhibit superior high-temperature mechanical properties over existing alloys [6]. Notably, unlike face-centered-cubic (fcc) MPEAs, bcc alloys do not exhibit radiation-hardening [26]. These findings are attributed to the distribution of point defects caused by high-energy ion bombardment. Therefore, it is of fundamental interest to understand the defect-related properties of MPEAs, such as vacancy stability and vacancy diffusion, especially their control.

Here we analyze defect properties of a refractory-based bcc MPEA, specifically we focus on $(Mo_{0.95}W_{0.05})_{0.85}Ta_{0.10}(TiZr)_{0.05}$ (denoted as MWTTZ), reported to exhibit 3× higher room-temperature elastic moduli [13] and superior oxidation resistance [27]. The high-temperature Young's modulus of MWTTZ (~2× at 2000 K relative to commercial alloys like TZM, composed of Ti-Zr-Mo) makes it an interesting alloy for in-depth analysis. The vacancy formation energy and migration energy barriers were calculated and their trends are presented. Our findings reveal that Ti has the lowest-energy migration barrier whereas W has the highest. We also find that lattice distortions caused by atomic displacements from their ideal alloy (x-ray average lattice) positions are directly proportional to the charge transfers between the neighboring atoms, impacting barrier energies. These revelations prove critical towards understanding the resistance to radiation damage of refractory-based bcc MPEAs.

## 2. Computational methods

*2.1 Density-Functional Theory (DFT):* We employed first-principles DFT as implemented in the Vienna *Ab-initio* Simulation Package (VASP) to investigate the effect of vacancies on structural and electronic properties of $(Mo_{0.95}W_{0.05})_{0.85}Ta_{0.10}(TiZr)_{0.05}$ [28, 29]. To represent the disordered MPEA configuration, we used a Super-Cell Random Approximate (SCRAPs) [33] optimized to get atomic point and pair correlation functions (3 to 5 neighbor shells) considering three different supercell sizes (80, 120, and 160 atoms) so as to determine the smallest (most computationally efficient) unit-cell size to still get reliable results. As expected, these sized SCRAPs had no effect on the vacancy formation energies ($E_{form}^{vac}$) [33]. For pure elements, 54-atom bcc (Mo/W/Ta) and hcp (Ti/Zr) supercells were used to simulate $E_{form}^{vac}$. A gamma-centered Monkhorst-Pack [30] k-mesh of $1 \times 2 \times 2$ ($3 \times 4 \times 4$) was used for bcc Brillouin-zone integration during geometry optimization (charge self-consistency). A kinetic-energy cutoff of 520 eV was used both for total energy ($10^{-5}$ eV/cell) and force (-0.001 eV/A) convergence. The Perdew, Burke and Ernzerhof (PBE) exchange-correlation for solids was used [31, 32].

*2.2 Vacancy-formation energy ($E_{form}^{Vac}$):* $E_{form}^{Vac}$ in the MPEA was estimated by

$$E_{form}^{Vac} = E_v - E_0 \pm \mu_v \quad , \quad \text{Eq. (1)}$$

where $E_v$ with vacancy and $E_0$ without vacancy, $\mu_v$ is the chemical potential of the defect species $v$, where the + sign (– sign) corresponds to a vacancy (interstitial) [34].

*2.3 Chemical potential ($\mu_v$):* The $\mu_v$ is defined as energy variation when one atom is added or removed from the system. In the solid-state, it is usually calculated as energy-per-atom of the corresponding element. Although the definition of $\mu_v$ remains the same in complex alloys, the values can be different from that of pure metals. However, in the literature, the variation of $\mu_v$ in pure metals (bulk) versus in an alloy were found to be insignificant [34, 35], as shown in Table 1. This suggests that, in most cases, taking elemental $\mu_v$ as starting point may save significant computational effort.

**Table 1.** Percentage chemical potential difference ($\Delta\mu_{Elements} = \mu_{alloy} - \mu_{metal}$) of alloying elements in alloys and pure metals [34-36].

| Sl. No. | Ref. | Metals | %$\Delta\mu$ |
|---|---|---|---|
| 1 | [34] | V, Ta, Cr, W | 1.28 |
| 2 | [35] | Cr, Mn, Fe, Co | 1.24 |
| 3 | [36] | Ni, Co, Cr, Fe | 3.04 |

Nevertheless, we reinvestigate the variation of $\mu_v$ in MPEAs [37] using relationship between elemental and alloy $\mu_v$ as

$$\mu_v + \sum_{i \neq v} c_i \mu_{eff}^v = \mu_{v/alloy}, \quad \text{Eq. (2)}$$

where $\mu_{v/alloy}$ is the chemical potential of metal $v$ when present in the alloy, $i$ refers to the site of the vacancy, and $\mu_{eff}^v$ can be calculated using $\mu_{eff}^v = \frac{\Delta E_0}{\Delta C_v}$, where $\Delta E_0 (= E_v - E_0)$ refers to the energy difference per atom between the systems with and without a vacancy and $\Delta C_v$ is the concentration change of the element $v$ with the finite-sized (N site) supercell in the vacancy and no vacancy cases ($\frac{1}{N-1} - \frac{1}{N} = \frac{1}{N(N-1)}$). The calculated values of $\mu_{eff}^v$ are listed in Table 2. Note that Eq. 2 reduces to $\mu_v = \mu_{v/alloy} = E_{alloy}/N$ for a pure metal, as $\sum_{i \neq v} c_i \mu_{eff}^v = 0$ when there is only a single species in the system. In Table 2, the difference between $\mu_v$ and $\mu_{v/alloy}$ is less than 1.1%, which matches recent reports; for instance, here $\mu_{v/alloy}$ for Ta is –11.70 eV-atom$^{-1}$ in agreement with –11.72 eV-atom$^{-1}$ [34]. Thus, $\mu_{v/alloy}$ in Table 2 was used throughout this work.

**Table 2.** Comparison of elemental chemical potentials ($\mu_v$) in pure metal and bcc MWTTZ.

| Element | Phase (VEC) | $\mu_v$ | $\mu_{v/alloy}$ |
|---|---|---|---|
|  |  | [eV-atom$^{-1}$] | |
| Ti | hcp (4) | -8.10 | -8.08 |
| Zr | hcp (4) | -8.52 | -8.51 |
| Ta | bcc (5) | -11.81 | -11.70 |
| Mo | bcc (6) | -10.80 | -10.71 |
| W | bcc (6) | -12.96 | -12.82 |

*2.4 Migration-energy barrier:* The energy barrier for defect migration was calculated using the climbing-image nudged-elastic band (cNEB) methodology [38]. The cNEB scheme generates the minimum energy path for the defect migration, which was calculated by generating multiple atomic configurations along a linear path that has negligible or no atomic forces normal to it. We used 6 intermediate atomic configurations to calculate the migration barrier. From the energies of each of those individual atomic configurations and subsequently from the energy profile of the migration path, the mechanism of the plausible defect migration and its energy barrier was calculated. This information is significant as it provides an insight into the kinetics of defect progression and also the mechanisms that possibly lead to defect recovery [39].

## 3. Results and discussions

*3.1 Vacancy-formation energies of pure metals:* Point defects are defined as imperfections in the crystal lattice with sizes of the order of the atomic diameter. The enthalpy of formation (E$_{form}$) is a characteristic of vacancies that governs the equilibrium concentration at any given temperature. The E$_{form}$ of interstitials is higher than that of the vacancy defects, therefore, vacancies naturally become dominate defect at equilibrium [40]. Experimentally, $E_{form}^{Vac}$ is the energy required to remove one atom from the interior of the crystal and replace it on the crystal surface [40]. The equilibrium concentration of point defects (both vacancy and interstitials) increases with temperature, which can be calculated by measuring specific heat, thermal expansion, electrical resistivity, positron annihilation, thermopower and perturbed angular correlation of γ–quanta [40]**.**

Information on causes of discrepancies between $E_{form}^{Vac}$ from various measurement methods relevant to this work is provided in the supplementary materials. In Table 3, we compare DFT calculated $E_{form}^{Vac}$ of metal atoms with experimental values (obtained via specific heat measurements and positron annihilation spectroscopy) [41]. This shows less than 6.5% difference for bcc elements, while a large deviation was found for hcp elemental metals. The observed discrepancy is attributed to experimental techniques, for example, the $E_{form}^{Vac}$ values via specific heat or by positron annihilation spectroscopy lead to completely different results [40].

**Table 3.** DFT-calculated $E_{form}^{Vac}$ in pure metals are compared with those obtained via specific heat or spectroscopy measurements [42-44]. A percent change was given by (Expt. −DFT) × 100/DFT.

| Metal atoms | $E_{form}^{vac}$ (eV/atom) | | % Change |
|---|---|---|---|
| | **DFT (this work)** | **Expt.** | |
| W | 3.26 | 3.30 [41] | 1.2 |
| Mo | 2.87 | 3.00 [41] | 4.3 |
| Ta | 2.90 | 3.10 [42] | 6.5 |
| Zr | 2.00 | 1.75 [43] | 14.3 |
| Ti | 2.07 | 1.55 [41] | 33.6 |

The $E_{form}^{Vac}$ for Ti in Table 3 has been compared to the experimental value that was obtained by specific heat measurements [41]. In the supplement, we have highlighted the possible ways that can lead to inaccuracy in $E_{form}^{Vac}$ during specific-heat measurements, i.e., the deviations with DFT $E_{form}^{Vac}$ can be attributed to type of measurements. To understand this further, we designed multiple vacancy configurations in a 54-atom supercell to calculate the $E_{form}^{Vac}$ to replicate the experimental scenario. We designed 6 atomic configurations with a 13.94 Å vacuum included in each configuration to avoid interactions among periodic images of both vacancy and surface-surface layers. Our strategy covers all possible scenarios of vacancy formation by removing atoms from 6 different positions in the supercell representing solid, as shown in Fig. S1. The vacancy created at surface layer by removing the atom to infinity in Fig. S1f most closely matches with the experiments. But this process does not reflect the mechanism of vacancy formation (removal of an interior atom from crystal associated with the appearance of an atom on crystal surface). The removal of an atom, one atom at a time from the surface to vacuum, is the representation of adatom binding energy [44]. Such an analysis is useful to study the energetic criteria for adsorption of an organic molecules on metal surfaces, which was recently used to facilitate user-defined vacancy patterning by the adsorption of $C_{60}$ molecules on gold, aluminum and beryllium surfaces [44] for applications in biosensors [45] and optoelectronics [46]. However, the $E_{form}^{Vac}$ calculated using the conventional approach is the main focus of this work, where an atom was extracted from the bulk to create the vacancy defects [34, 35, 47].

*3.2 MPEA vacancy-formation energies*: $E_{form}^{Vac}$ of each element in MWTTZ are listed in Table 2. The elemental chemical potentials in Table 2 were used to estimate $E_{form}^{Vac}$ using Eq. 1. In Table 4, $E_{form}^{Vac}$ shows a wide range from 2.87 eV for Zr to 3.84 eV for Mo, similar to other reports [36, 48]. The energy spread of 0.97 eV can potentially be attributed to the considerable local lattice distortions. The $E_{form}^{Vac,alloy}$ shows increase with respect to $E_{form}^{Vac,v}$ with variation from 7% to 74%, where W shows an increase of 7% while Ti is 74%. Tungsten with highest melting point (3422°C) undergoes the least change in $E_{form}^{Vac}$, which comes from stronger metallic bonding as W has a larger valence-electron count (VEC=6) than Ti/Zr (VEC=4) and Ta (VEC=5). A similar effect was observed for Mo (VEC=6). On the other hand, Ti shows maximum increase in $E_{form}^{Vac}$ compared to the hcp Ti, where the metallic bonds are weaker, as reflected in its lowest melting point of 1668°C. We also show a chemical environment dependence of $E_{form}^{Vac}$ in Table 4, where W shows 6% drop compared to the pure metal, while all other elements show significant increase.

**Table 4.** $E_{form}^{Vac}$ of each alloying element of MWTTZ compared to pure-metal value, with percent change calculated as $[E_{form,alloy}^{Vac} - E_{form,v}^{Vac}] \times 100/E_{form,v}^{Vac}$.

| Element | $E_{form,alloy}^{Vac}$ [eV] | | $E_{form,v}^{Vac}$ [eV] | % Change | |
|---|---|---|---|---|---|
| | Env#1 | Env#2 | Pure metal | Env#1 | Env#2 |
| W | 3.49 | 3.07 | 3.26 | 7.1 | –5.8 |
| Ta | 3.79 | 3.71 | 2.90 | 30.7 | 27.9 |
| Mo | 3.84 | 3.57 | 2.87 | 33.8 | 24.4 |
| Zr | 2.87 | 3.23 | 2.00 | 43.5 | 61.5 |
| Ti | 3.62 | 3.42 | 2.07 | 74.9 | 65.2 |

Thermodynamic properties depend strongly on the local chemical environment and atomic interaction [33], even more critical for non-stoichiometric cases as the unit-cell are more complex depending on composition. Therefore, it is wise to explore the effect of SCRAP size on $E_{form}^{Vac}$ of MWTTZ before going into more detail. In Fig. 1, we plot $E_{form}^{Vac}$ for each alloying element calculated in three different sizes of SCRAPs with 80, 120 and 160 atoms. Our calculations show a considerable increase in $E_{form}^{Vac}$ from 0.1 to 0.5 eV/atom, however, these effects diminish for 120 and 160 atom SCRAPs. This suggests that careful evaluation of size effect is important in MPEAs, especially for non-stoichiometric compositions. More importantly, having cells that are large enough are critical to avoid interaction of vacancy or defects with its periodic images; therefore, SCRAP sizes were such that the image distance between two vacancies remains at a minimum of 12Å. Hence, a 160-atom SCRAP was used in all vacancy-related calculations to avoid numerical issues and size-dependent effects.

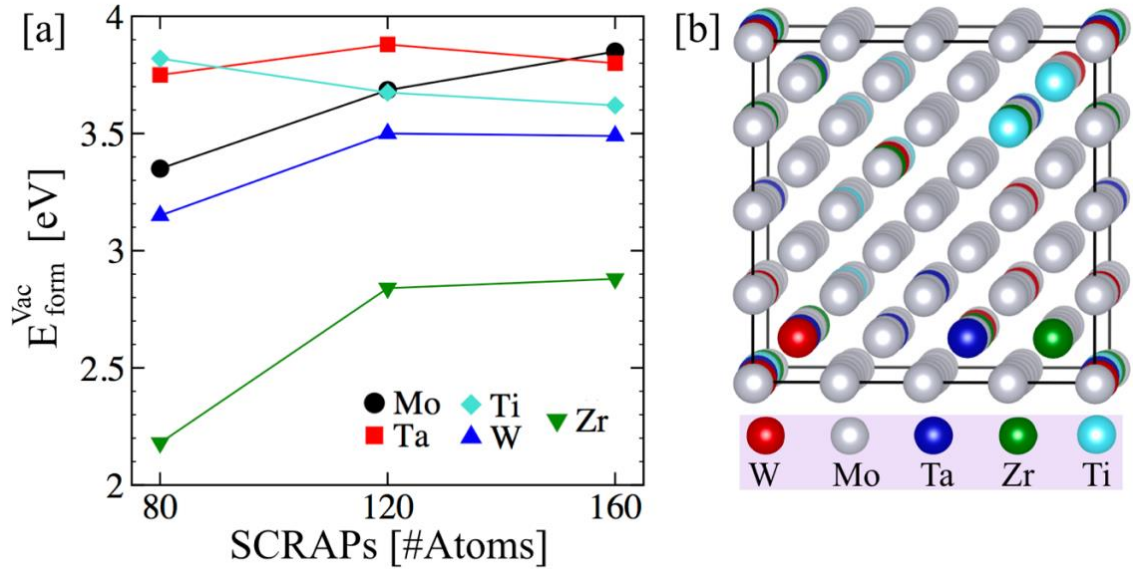

**Figure 1.** In MWTTZ, (a) dependence on SCRAP size for $E_{form}^{Vac}$ for each element. A considerable change in $E_{form}^{Vac}$ was found for W/Mo/Zr with increasing cell size. (b) Schematic of 160-atom (4x4x5x2-atom bcc cell) SCRAP used in this work for off-stoichiometric MPEAs.

*3.3 Local Lattice Distortions*

Solid-solution strengthening lead to improved mechanical properties of refractory MPEAs and is often attributed to local lattice distortions (LLD) driven by chemical complexity of alloying [49-51]. However, we find no reports on how point-defects impact LLD and any mechanism that can control this behavior. Figure 2 shows the vector displacements of the first-nearest neighbors (NN) in environment 1 (Fig. 2a,b) and second-NN in environment 2 (Fig. 2c,d) for a Ti vacancy compared to no-vacancy case. Environment 1 and 2 indicate different positions in the lattice where a specific metal vacancy was created by removing a metal atom. A Zr atom in the first NN in both the environments undergoes a large distortion in the presence of a vacancy, which occurs due to a charge deficiency (vacancy) created by the missing Ti atom and, as such, the Zr atom readjust itself with the neighboring atoms to compensate for that deficiency. Atomic displacements near vacancy sites are larger due to need of increased charge readjustment caused by the vacancy. The observation that unlike-atom species in the NNs cause higher displacements of NNs is consistent through all metal vacancies. The vector atomic displacements from Mo/Ta/W/Zr vacancy are shown in the supplementary Fig. S2-S5.

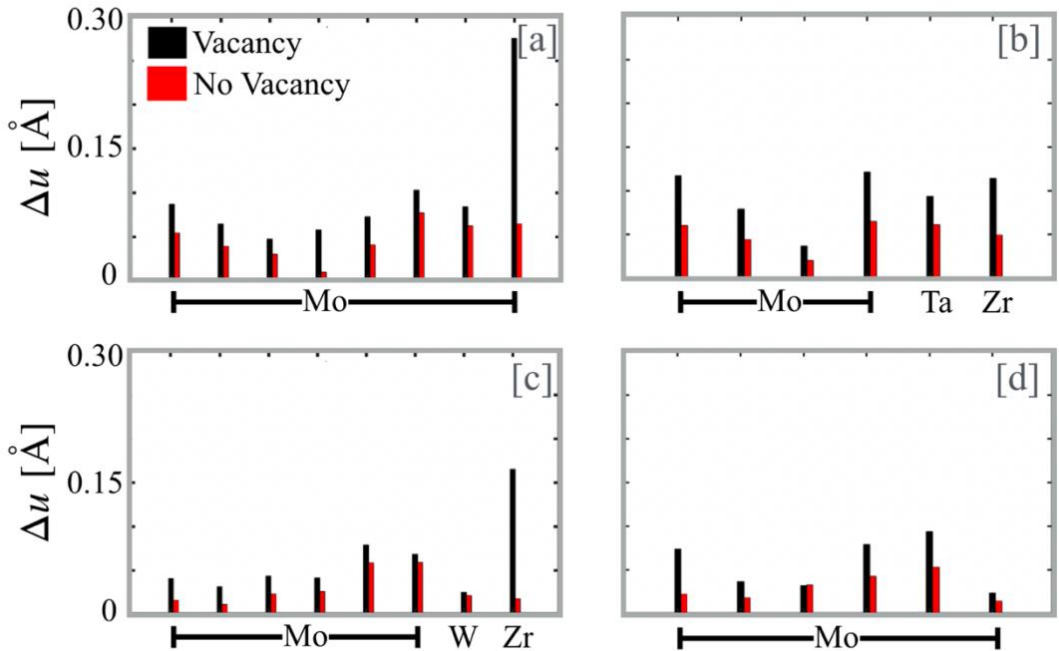

**Figure 2.** Vector atomic displacements in MWTTZ of (a, c) first-, and (b, d) second-nearest neighbor environment around a Ti vacancy (black bars) compared to no-vacancy case (red bars).

Recently, Song et al. [49] have shown that hard-sphere model (δ size-difference parameter [52]) fails to estimate the LLD accurately, e.g., it overestimates in refractory MPEAs and underestimates in 3$d$ MPEAs. So, to avoid using empirical estimates, we explored the quantum-mechanical origin of LLD through DFT charge density. Atoms in complex MPEAs with diverse chemical environment can lead to varying degree of interaction resulting from strong charge fluctuation among unlike-metal-atom pairs with differing electronegativities. We hypothesize a significant charge transfer between unlike species, resulting in larger atomic displacements from their ideal lattice (average x-ray) positions. In Fig. 3, charge-density differences ($\Delta\rho = \rho_{vac}$- $\rho_{no-vac}$) are plotted with and without vacancy, where charge readjustments of no-vacancy case (colored blue) are compared to vacancy cases (colored yellow) at various metal sites.

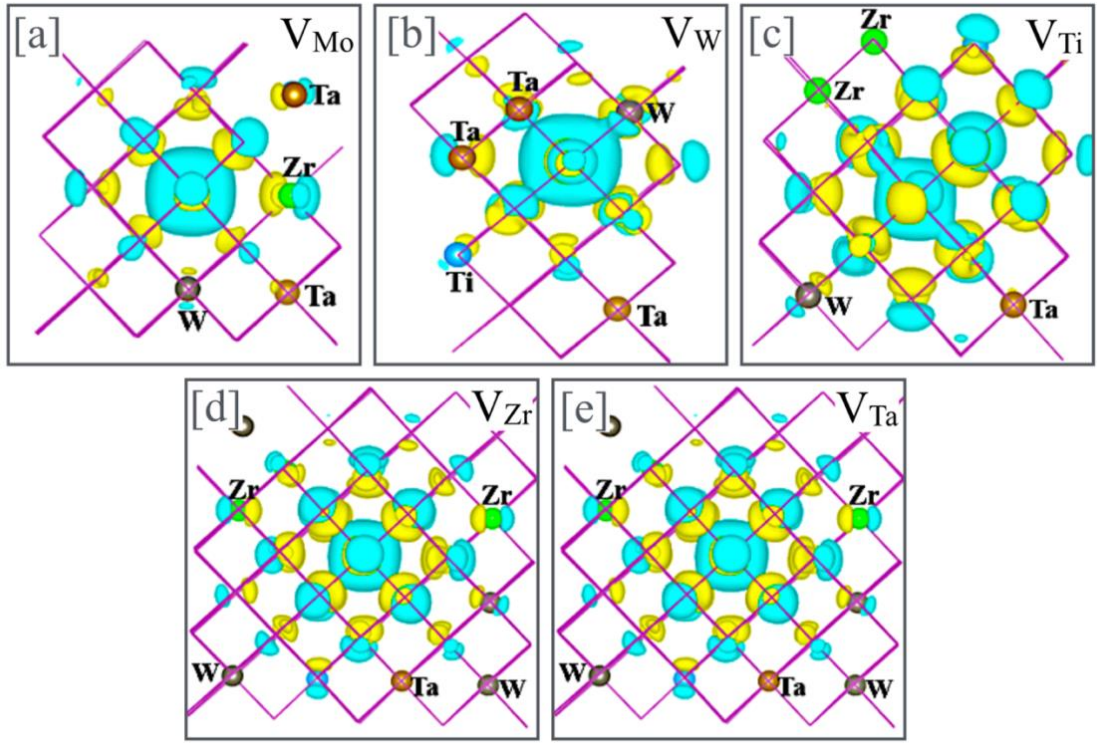

**Figure 3.** Charge-density difference ($\Delta \rho = \rho_{vac} - \rho_{no-vac}$) in MWTTZ between no-vacancy and vacancy cases of (a) $V_{Mo}$, (b) $V_W$, (c) $V_{Ti}$, (d) $V_{Zr}$ and (e) $V_{Ta}$. Effect of $V_{Zr}$ on charge redistribution in (c) was strongest (out to 5$^{th}$ NNs) and correlated with Zr being the least electronegative among all elements.

In Fig. 3a, the charge redistribution is more localized (2$^{nd}$ NNs) for Mo vacancy, so the LLD does not show much effect on environment in MWTTZ. So, a Mo vacancy creates no significant charge redistribution due to weaker effect of charge-transfer with 1$^{st}$ and 2$^{nd}$ NNs. As VEC of Mo/W are the same, and with similar atomic sizes, the same effect was observed for W (Fig. 3b), so Mo/W are expected to have same effect on LLD. However, Ti/Zr shows local lattice relaxation effect up to 4$^{th}$ to 5$^{th}$ NNs due to charge redistribution. This occurs likely due to large charge-variation introduced by Ti/Zr vacancies, as is obvious in Fig. 3c,d, where $\Delta \rho$ is beyond 2$^{nd}$ NNs [34, 53]. The Ta (group V, VEC=5) vacancy in Fig. 3e shows behavior intermediate to Mo/W (group VI, VEC=5) and Ti/Zr (group IV, VEC=4) as its VEC is intermediate. This behavior is explained via electronegativity ($\chi$) difference, where atoms with higher $\chi$ should pull more charge than atoms with lower $\chi$. If Fig. 3a,e is analyzed based on electronegativities, then Zr ($\chi$ =1.33), Ti ($\chi$ =1.54) and Ta ($\chi$ =1.50) in a mostly Mo-rich environment show more delocalized charges from $3^{rd}$ to $5^{th}$ NNs while elements with higher electronegativity, i.e., Mo ($\chi$=2.16) < W ($\chi$=2.36), show more localized charges due to higher electron affinity.

In Fig. 3, a vacancy causes a varying degree of electron transfer depending on atom type, so a varying strength of metallic bonds should be expected. The qualitative quantum-mechanical insights in Fig. 3 shows the mechanism controlling atomic displacements before and after creation of a vacancy. Such a mechanism has been previously noted to induce LLD [34] [47], but we provide more quantitative inference in Fig. 4 by plotting $1^{st}$-NN atom net-displacement versus charge transfer. The net displacement is defined as the difference with respect to ideal bcc as $\sqrt{(x1-x)^2 + (y1-y)^2 + (z1-z)^2}$ (with no-vacancy lattices (x, y, z) vs. vacancy (x1, y1, z1)), and $\Delta\rho$ is the difference between vacancy and no-vacancy cases.

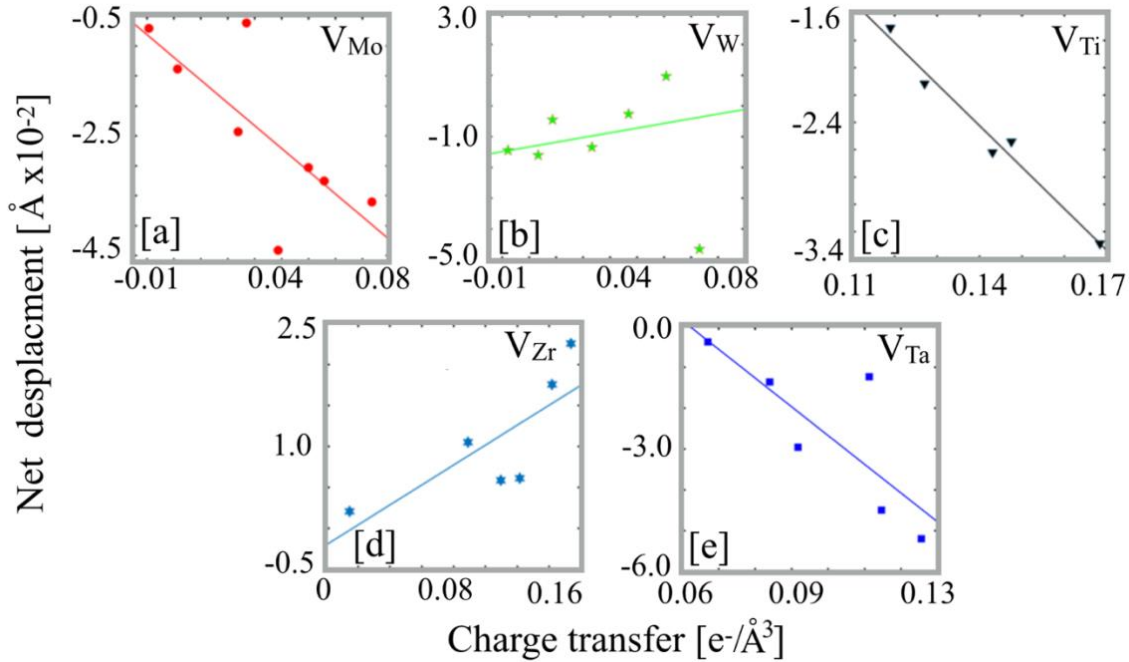

**Figure 4.** Correlation between charge-transfer ($e^-/Å^3$) and net-displacement in MWTTZ at Mo sites in the first-neighbors for (a) $V_{Mo}$, (b) $V_W$, (c) $V_{Ti}$, (d) $V_{Zr}$ and (e) $V_{Ta}$ vacancies with respect to no-vacancy case. For simplicity, only Mo NNs around the vacancy sites were considered in the charge analysis.

As discussed in Fig. 3, charges redistribution is explained through the ordering of elemental electronegativity. The quantitative analysis of charge transfer in Fig. 4a,e shows that Ti (1.54), Zr (1.33) and Ta (1.5) tend to undergo a maximum redistribution of charges in the range 0.13-0.17 $e^-/Å^3$. In contrast, Mo (2.16)/W(2.36) with higher electronegativity shows charge transfer in the similar range, i.e., 0.08 e⁻/Å³. Despite great variance, each metal vacancy shows a near linear trend of atomic displacements versus charge transfer where Ti, Zr and Ta show highest displacements owing to the maximum degree of charge transfer.

In the 160-atom SCRAP (128 Mo sites), inclusion of other neighbors than Mo may result in more outliers owing to other atomic properties like atomic mass and radius in charge analysis. Therefore, for consistency, the displacement versus charge-transfer plotted in Fig. 4a,e includes only Mo atoms as the 1$^{st}$ NN. We observed that electronic rearrangement increased the metallic character around vacancy sites, where sites with large charge fluctuation show increased atomic displacement from ideal sites that directly relates to LLD. The increased LLD makes it difficult to move or extract an atom out of the lattice and increases $E_{form}^{Vac}$, as shown in Table 4.

The presence of certain outliers in Fig. 4a,e is conjectured to show up in cases where the 2$^{nd}$ NNs has some role in charge transfer. For instance, for the W vacancy in Fig. 4b, the displacement of one of the surrounding Mo atoms breaks from linearity. On the other hand, the presence of Zr in the 2$^{nd}$ NNs (see Fig. S5) displaces the Mo atom due to a large electronegativity difference between the Zr (1.33) and Mo (2.16). Similarly, Zr in the 2$^{nd}$ NNs in Mo vacancy case shows few outliers.

*3.4 Vacancy-migration energies*: Vacancies are the most common type of point defects at equilibrium and they largely control the diffusion kinetics. At elevated temperatures, vacancy diffusion becomes a critical aspect as the vacancy migration controls the chemistry driven transformations, for example corrosion and creep [54]. To date, few studies have been done to explore the vacancy-diffusion mechanism in refractory-based bcc MPEAs [34]. In Fig. 5a-e, we analyzed vacancy migration barriers in disordered MWTTZ and compare to those of pure metals.

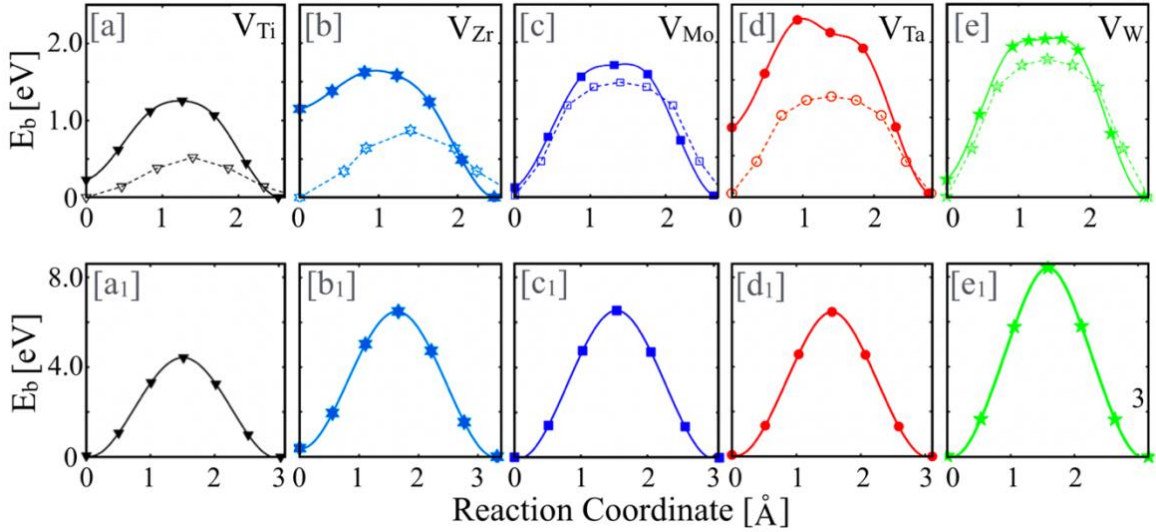

**Figure 5.** Calculated vacancy-migration energy and barriers ($E_b$) for MWTTZ along (a-e) $\frac{1}{2}\langle 111\rangle$ direction (solid lines) compared to barriers of pure metals (dashed lines), and ($a_1$-$e_1$) along $\langle 100\rangle$ direction. Pure metal $E_b$ are from Ref. [55] ($V_{Ti}$), Ref. [56] ($V_{Zr}$), and Ref. [57] ($V_{Mo}$, $V_{Ta}$, and $V_W$). Migration energies appear to increase with group number from periodic table.

We used cNEB to calculate migration barriers in MWTTZ. cNEB generates a sequence of atomic configurations and provides an information for warping of the migration path (i.e., "elastic band") with a goal to move towards a low-energy path in the landscape of potential-energy surface. The diffusion barriers for vacancies in the alloy versus pure metals are shown in Fig. 5a-e, where energy barriers in alloy matrix (solid lines) are compared to those in pure metals (dashed lines) [51-53]. The energy barrier in the alloy for Ti/Zr shows significant increase. This was explained through the charge-density readjustment (electronegativities) caused by vacancies in different chemical environments (Fig. 3) that leads to significant increase in $E_{form}^{Vac}$ of Ti/Zr compared to Mo/W/Ta, as listed in Table 3. We attribute this fact to the larger readjustment of charges up to 5$^{th}$-nearest neighbors for hcp metals vacancy (Ti/Zr) compared to bcc metal vacancies, i.e., increased interaction among the Zr neighbors. Interestingly, Ti/Zr also show anisotropic charge distribution that points out towards strong environment dependence, also visible in migration barrier of Zr in Fig. 5b. This phenomenon can also be understood in terms of pair distribution function that shows increased interatomic interactions between unlike atomic pairs in Fig. S6 as compared to the interactions between like pairs in MWTTZ. Similar behavior has been reported previously [15], where the pair-correlation function [$g_{\alpha\beta}(r)$] of unlike pairs was found to have a higher interaction than like pairs. For example, Ta-Ta pairs in Fig. S6 are weaker than the Ta-Mo,

Ta-Zr and Ta-W, which further affirms our hypothesis. Hence, when metal elements crystallize in an alloy, the migration of an atom to a neighboring vacancy site becomes formidable due to the strong interaction arising from the diverse environment, which depends on type of neighbor as well as diversity of neighbors.

The magnitude of electronegativity ($\chi$) represents the ability of element to attract the charge, i.e., higher $\chi$ indicates higher affinity to pull charge. In the increasing order, $\chi$ of each alloying elements is Zr (1.33) < Ta (1.50) < Ti (1.54) < Mo (2.16) < W (2.36). Higher $\chi$ would mean a larger $\Delta\chi$ and larger charge exchange with NNs, therefore, a stronger interaction. This makes the complex concentrated alloy unfavorable for vacancy diffusion. Hence, we anticipate that the order of migration barriers should be proportional to the $\chi$ of elements. If we closely look at migration barriers in Fig. 5, indeed, the migration barrier of each element falls in the same order as their $\chi$ with the exception of Ti. A lower atomic mass of Ti may override the $\chi$ contribution, where the lowest atomic mass allows a higher vibrational frequency and velocity at a given temperature compared to other elements, perhaps playing a role in lowering Ti diffusion barrier.

We also investigated the directional dependence of migration barrier in MWTTZ. The $\frac{1}{2}$<111> Burger's vector in [110] plane in Fig. 5a-e shows 2-3 times lower barriers compared to <100> in [100] plane in Fig. 5a$_1$-e$_1$. Notably, the <111> direction lies in close-packed [110] bcc plane, which has higher packing fraction (higher atomic density). Therefore, the energy required to move vacancy on a close-packed plane under external force will be relatively smaller than the less closed-packed [100] direction, where the atoms have lower atomic density.

Diffusion energetics in Fig. 5 shows that Ti is relativity favorable compared to Zr/Ta and Mo/W and will resist the diffusion. Therefore, to emphasize on element specific diffusion, we discuss the environment dependence of migration barrier in MWTTZ. The barrier heights for all metals in MPEA show dependence on local environment, especially Zr and Mo. We discuss the case of Zr (Fig. 6) as it shows the maximum variation in diffusion barrier with change in local environment. Figure 6a and 6b show environment and directional dependence, respectively. We observed a considerable disparity in migration barrier depending upon the local environment (Fig. 6a), where $\Delta E_b^{Zr1}$ = 0.48 eV in environment 1 while $\Delta E_b^{Zr2}$ = 1.63 eV in environment 2. Experimentally, a higher diffusion barrier for Zr in environment 2 (with its larger charge redistribution near Zr-vacancy site than in environment 1) indicates that vacancies are difficult to move, i.e., a significant fraction of the vacancies, once created, will remain intact during the irradiation process, but those in environment 1 should more easily diffuse.

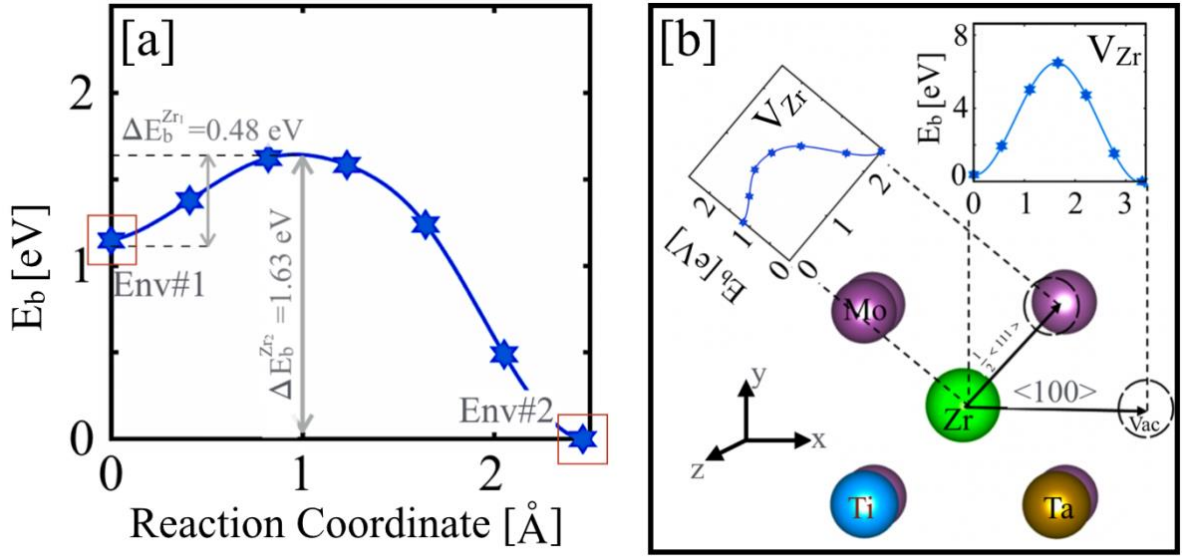

**Figure 6.** (a) Environmental and (b) directional dependence of migration energy barrier for $V_{Zr}$ in MWTTZ. Migration barrier in <100> is higher by a factor of 2-3 over $\frac{1}{2}$<111> due to lower close-packing in <100> plane, i.e., low linear density. A significant variation in migration barrier (environment 1 $\Delta E_b^{Zr1}$ = 0.48 eV and environment 2 $\Delta E_b^{Zr2}$ = 1.63 eV) suggests that vacancies created in environment 2 may remain intact through the irradiation exposure period but those in environment 1 are more likely to undergo diffusion.

**Jump Frequency:** Here we present a mean-field-like understanding of one-dimensional jump frequency of constituent elements of MWTTZ. The defect jump rate can be described in terms of the Arrhenius relation [59,60], which can be defined as

$$\Gamma = \nu e^{-E_b/k_B T}, \qquad \text{Eq. (3)}$$

where $E_b$ is the energy barrier required to move defect to the saddle point from an equilibrium position, $k_B$ is Boltzmann's constant, T is absolute temperature (Kelvin), and $\nu$ is the vibrational frequency of defect along the saddle point. The results in **Fig. 6** show that $\Gamma$ in the disorder MWTTZ is strongly correlated with the migration barrier, as shown in **Fig. 5**, where the order of jump rate of defect at each defect site to jump back to its original position is $\Gamma(W) > \Gamma(Mo) > \Gamma(Ti) > \Gamma(Ta) > \Gamma(Zr)$. This important finding indicates that a jumping atom to defect site are surrounded by other atoms with which it interacts, and which can greatly impact the diffusion of atom in solid-solution phase. The defect interaction with different atoms in varying degree of environment in MPEAs are greatly impacted by the probability of defects jump.

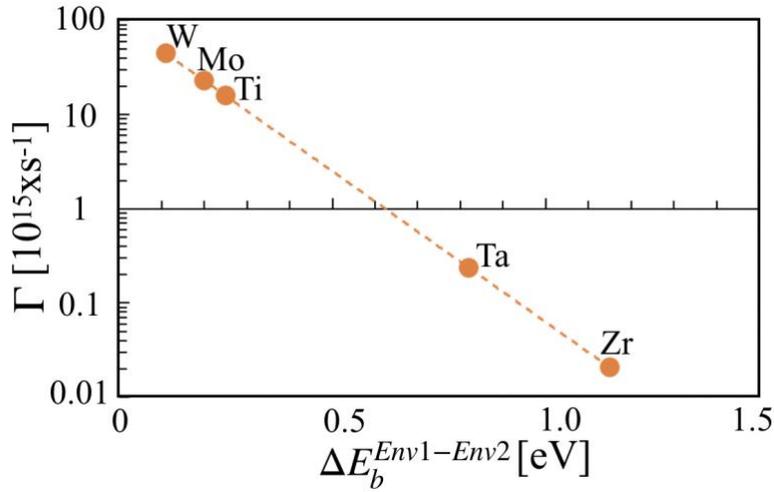

**Figure 6.** Defect jump-rate (Γ) plot on logarithmic scale in MWTTZ. The order of jump-rate shows strong correlation with difference of diffusion barrier of defects in two different environments in **Fig. 5**.

## 4. Conclusions

We investigated the environment dependence of the vacancy-formation energy $E_{form}^{Vac}$ and associated diffusion barrier in $(Mo_{0.95}W_{0.05})_{0.85}Ta_{0.10}(TiZr)_{0.05}$, a refractory-based bcc MPEA. The $E_{form}^{Vac}$ of metal atoms in this MPEA was found to increase from 7% for W to 74% for Ti compared to the pure metals. The higher $E_{form}^{Vac}$ of Ti suggests that a Ti vacancy is hard to create, however, a low diffusion barrier for Ti indicates that once Ti vacancy is formed, the migration becomes relatively easier compared to other atom types. The charge-transfer effect (governed by electronegativities) establishes that local lattice distortions (LLD) in MPEAs has a quantum origin that can be tuned through controlled variation of local atomic environments. We exemplified this with DFT atomic displacements versus local environments with and without a vacancy and showed displacements greatly vary with type of metal-vacancy and its local environments. Indeed, the atomic displacement showed approximately linear dependence on the net charge transfer.

      We also calculated migration-energy barrier ($E_b^{Vac}$) for atoms in this refractory MPEA. The $E_b^{Vac}$ are significantly higher for Ti and Mo than pure-metal state, while other metal atoms very weak dependence. Importantly, the large $E_b^{Vac}$ in the MPEA was attributed to stronger unlike pair interactions that make migration difficult. $E_b^{Vac}$ also shows direct correlation with electronegativity, where higher electronegative atoms show higher barriers. The migration-energy barriers show a strong environmental and directional dependence. For example, the barriers for

Zr $\Delta E_b^{Zr}$ varied by a factor of 3 (0.48 eV to 1.63 eV) depending on local starting environment; and, for directional dependence, Zr shows a factor of 2-3 higher $E_b^{Vac}$ along <100> direction than $\frac{1}{2}$<111> direction due to lower packing fraction making diffusion difficult in bcc alloys. Atomic displacements provided a strong correlation between charge redistribution, LLD, and environment for unique insight into point-defect properties in refractory MPEAs. For defect formation and migration, this study provided chemical insights to manipulate and improve the high-temperature thermo-mechanical properties of refractory MPEAs, especially for use in extreme environments, affected by vacancy formation and diffusion, like creep.

**Acknowledgements**

Ankit Roy, while visiting Ames Laboratory, was supported in part by a fellowship from the National Science Foundation (NSF) through award CMMI-1944040. General theory developments at Ames Laboratory were supported by the Department of Energy (DOE), Office of Science, Basic Energy Sciences, Materials Science and Engineering Department. Ames Laboratory is operated by Iowa State University for the U.S. DOE under contract DE-AC02-07CH11358. Specific theory application to MWTTZ was supported by the U.S. DOE, Office of Energy Efficiency and Renewable Energy, Advanced Manufacturing Office (AMO) project WBS 2.1.0.19.

**CRediT authorship contribution statement**
**Ankit Roy**: DFT calculations, Analysis, Writing – original draft, Writing – review & editing. **Prashant Singh**: Conceptualization, Supervision, Analysis, Writing – original draft, Writing – review & editing. **Ganesh Balasubramanian**: Funding acquisition, Supervision, Resources, Writing – review & editing. **Duane D. Johnson**: Conceptualization, Analysis, Funding acquisition, Supervision, Resources, Writing – review & editing.

**Data availability**

The authors declare that the data supporting the findings of this study are available within the paper and supplement. Also, the data that support the plots within this paper and other finding of this study are available from the corresponding author upon reasonable request.

**Declaration of Competing Interest**

The authors declare that they have no known competing financial interests or personal relationships that could have appeared to influence the work reported in this paper.

# References

[1] S. Carnot, Reflections on the motive power of heat and on machines fitted to develop that power, J. Wiley, 1890.

[2] J.W. Yeh, S.K. Chen, S.J. Lin, J.Y. Gan, T.S. Chin, T.T. Shun, C.H. Tsau, S.Y. Chang, Nanostructured high-entropy alloys with multiple principal elements: novel alloy design concepts and outcomes, Advanced Engineering Materials, 6 (2004) 299-303.

[3] B. Cantor, I.T.H. Chang, P. Knight, A.J.B. Vincent, Microstructural development in equiatomic multicomponent alloys, Mat Sci Eng a-Struct, 375 (2004) 213-218.

[4] E.P. George, W. Curtin, C.C. Tasan, High entropy alloys: A focused review of mechanical properties and deformation mechanisms, Acta Materialia, 188 (2020) 435-474.

[5] D.B. Miracle, O.N. Senkov, A critical review of high entropy alloys and related concepts, Acta Materialia, 122 (2017) 448-511.

[6] O.N. Senkov, G.B. Wilks, J.M. Scott, D.B. Miracle, Mechanical properties of Nb25Mo25Ta25W25 and V20Nb20Mo20Ta20W20 refractory high entropy alloys, Intermetallics, 19 (2011) 698-706.

[7] A. Roy, T. Babuska, B. Krick, G. Balasubramanian, Machine learned feature identification for predicting phase and Young's modulus of low-, medium-and high-entropy alloys, Scripta Materialia, 185 (2020) 152-158.

[8] J.M. Rickman, H.M. Chan, M.P. Harmer, J.A. Smeltzer, C.J. Marvel, A. Roy, G. Balasubramanian, Materials informatics for the screening of multi-principal elements and high-entropy alloys, Nat Commun, 10 (2019) 2618.

[9] M.-H. Tsai, J.-W. Yeh, High-entropy alloys: a critical review, Materials Research Letters, 2 (2014) 107-123.

[10] A. Roy, P. Sreeramagiri, T. Babuska, B. Krick, P.K. Ray, G. Balasubramanian, Lattice distortion as an estimator of solid solution strengthening in high-entropy alloys, Materials Characterization, (2021) 110877.

[11] Z. Wu, H. Bei, G.M. Pharr, E.P. George, Temperature dependence of the mechanical properties of equiatomic solid solution alloys with face-centered cubic crystal structures, Acta Materialia, 81 (2014) 428-441.

[12] M. Gianelle, A. Kundu, K. Anderson, A. Roy, G. Balasubramanian, H.M. Chan, A novel ceramic derived processing route for Multi-Principal Element Alloys, Materials Science and Engineering: A, 793 (2020) 139892.

[13] P. Singh, A. Sharma, A.V. Smirnov, M.S. Diallo, P.K. Ray, G. Balasubramanian, D.D. Johnson, Design of high-strength refractory complex solid-solution alloys, npj Computational Materials, 4 (2018) 16.

[14] D. Miracle, High entropy alloys as a bold step forward in alloy development, Nature communications, 10 (2019) 1-3.

[15] A. Roy, J. Munshi, G. Balasubramanian, Low energy atomic traps sluggardize the diffusion in compositionally complex refractory alloys, Intermetallics, 131 (2021) 107106.

[16] Q. Ding, Y. Zhang, X. Chen, X. Fu, D. Chen, S. Chen, L. Gu, F. Wei, H. Bei, Y. Gao, Tuning element distribution, structure and properties by composition in high-entropy alloys, Nature, 574 (2019) 223-227.

[17] P. Sreeramagiri, Roy, A. & Balasubramanian, G. , Effect of Cooling Rate on the Phase Formation of AlCoCrFeNi High-Entropy Alloy., J. Phase Equilib. Diffus. (2021). (2021).

[18] A. Roy, G. Balasubramanian, Predictive descriptors in machine learning and data-enabled explorations of high-entropy alloys, Computational Materials Science, (2021) 110381.

[19] T. Kostiuchenko, F. Körmann, J. Neugebauer, A. Shapeev, Impact of lattice relaxations on phase transitions in a high-entropy alloy studied by machine-learning potentials, npj Computational Materials, 5 (2019) 1-7.

[20] S.J. Zinkle, J.T. Busby, Structural materials for fission & fusion energy, Materials today, 12 (2009) 12-19.

[21] V. Philipps, Tungsten as material for plasma-facing components in fusion devices, Journal of nuclear materials, 415 (2011) S2-S9.
[22] J. Marian, C.S. Becquart, C. Domain, S.L. Dudarev, M.R. Gilbert, R.J. Kurtz, D.R. Mason, K. Nordlund, A.E. Sand, L.L. Snead, Recent advances in modeling and simulation of the exposure and response of tungsten to fusion energy conditions, Nuclear Fusion, 57 (2017) 092008.
[23] O. El-Atwani, N. Li, M. Li, A. Devaraj, J. Baldwin, M.M. Schneider, D. Sobieraj, J.S. Wróbel, D. Nguyen-Manh, S.A. Maloy, Outstanding radiation resistance of tungsten-based high-entropy alloys, Science advances, 5 (2019) eaav2002.
[24] C. Becquart, C. Domain, Modeling microstructure and irradiation effects, Metallurgical and Materials Transactions A, 42 (2011) 852-870.
[25] T. Suzudo, M. Yamaguchi, A. Hasegawa, Stability and mobility of rhenium and osmium in tungsten: first principles study, Modelling and Simulation in Materials Science and Engineering, 22 (2014) 075006.
[26] Y. Lu, H. Huang, X. Gao, C. Ren, J. Gao, H. Zhang, S. Zheng, Q. Jin, Y. Zhao, C. Lu, A promising new class of irradiation tolerant materials: Ti2ZrHfV0. 5Mo0. 2 high-entropy alloy, Journal of materials science & technology, 35 (2019) 369-373.
[27] R. Su, H. Zhang, G. Ouyang, L. Liu, W. Nachlas, J. Cui, D.D. Johnson, J.H. Perepezko, Enhanced oxidation resistance of (Mo95W5) 85Ta10 (TiZr) 5 refractory multi-principal element alloy up to 1300° C, Acta Materialia, 215 (2021) 117114.
[28] G. Kresse, D. Joubert, From ultrasoft pseudopotentials to the projector augmented-wave method, Physical review b, 59 (1999) 1758.
[29] G. Kresse, J. Hafner, Ab initio molecular dynamics for liquid metals, Physical Review B, 47 (1993) 558.
[30] H.J. Monkhorst, J.D. Pack, Special points for Brillouin-zone integrations, Physical review B, 13 (1976) 5188.
[31] P.E. Blöchl, Projector augmented-wave method, Physical review B, 50 (1994) 17953.
[32] J.P. Perdew, K. Burke, M. Ernzerhof, Generalized gradient approximation made simple, Physical review letters, 77 (1996) 3865.
[33] R. Singh, A. Sharma, P. Singh, G. Balasubramanian, D.D. Johnson, Accelerating computational modeling and design of high-entropy alloys, Nature Computational Science, 1 (2021) 54-61.
[34] S. Zhao, Defect properties in a VTaCrW equiatomic high entropy alloy (HEA) with the body centered cubic (bcc) structure, Journal of Materials Science & Technology, 44 (2020) 133-139.
[35] M. Mizuno, K. Sugita, H. Araki, Defect energetics for diffusion in CrMnFeCoNi high-entropy alloy from first-principles calculations, Computational Materials Science, 170 (2019) 109163.
[36] S. Zhao, T. Egami, G.M. Stocks, Y. Zhang, Effect of d electrons on defect properties in equiatomic NiCoCr and NiCoFeCr concentrated solid solution alloys, Physical Review Materials, 2 (2018) 013602.
[37] A. Esfandiarpour, M. Nasrabadi, Vacancy formation energy in CuNiCo equimolar alloy and CuNiCoFe high entropy alloy: ab initio based study, Calphad, 66 (2019) 101634.
[38] G. Henkelman, B.P. Uberuaga, H. Jónsson, A climbing image nudged elastic band method for finding saddle points and minimum energy paths, Journal of chemical physics, 113 (2000) 9901-9904.
[39] S. Middleburgh, R. Grimes, Defects and transport processes in beryllium, Acta materialia, 59 (2011) 7095-7103.
[40] Y. Kraftmakher, Equilibrium vacancies and thermophysical properties of metals, Physics Reports, 299 (1998) 79-188.
[41] B. Medasani, M. Haranczyk, A. Canning, M. Asta, Vacancy formation energies in metals: A comparison of MetaGGA with LDA and GGA exchange–correlation functionals, Computational Materials Science, 101 (2015) 96-107.
[42] P. Jung, Production of atomic defects in metals, Landolt-Börnstein, Croup III: Crystal and Solid State Physics, Ed. H. Ullmaier, Springer-Verlag, Berlin, 25 (1991).

[43] G. Hood, R. Schultz, J. Jackman, Recovery of single crystal. cap alpha.-Zr from low temperature electron irradiation. A positron annihilation spectroscopy study, J. Nucl. Mater.;(Netherlands), 126 (1984).
[44] A. Kaiser, F. Vines, F. Illas, M. Ritter, F. Hagelberg, M. Probst, Vacancy patterning and patterning vacancies: controlled self-assembly of fullerenes on metal surfaces, Nanoscale, 6 (2014) 10850-10858.
[45] K. Schierbaum, T. Weiss, E.T. Van Veizen, J. Engbersen, D. Reinhoudt, W. Göpel, Molecular recognition by self-assembled monolayers of cavitand receptors, Science, 265 (1994) 1413-1415.
[46] E. Menard, M.A. Meitl, Y. Sun, J.-U. Park, D.J.-L. Shir, Y.-S. Nam, S. Jeon, J.A. Rogers, Micro-and nanopatterning techniques for organic electronic and optoelectronic systems, Chemical reviews, 107 (2007) 1117-1160.
[47] P. Singh, D. Sauceda, R. Arroyave, The effect of chemical disorder on defect formation and migration in disordered max phases, Acta Materialia, 184 (2020) 50-58.
[48] C. Li, J. Yin, K. Odbadrakh, B.C. Sales, S.J. Zinkle, G.M. Stocks, B.D. Wirth, First principle study of magnetism and vacancy energetics in a near equimolar NiFeMnCr high entropy alloy, Journal of Applied Physics, 125 (2019) 155103.
[49] H. Song, F. Tian, Q.-M. Hu, L. Vitos, Y. Wang, J. Shen, N. Chen, Local lattice distortion in high-entropy alloys, Physical Review Materials, 1 (2017) 023404.
[50] Y. Ikeda, K. Gubaev, J. Neugebauer, B. Grabowski, F. Körmann, Chemically induced local lattice distortions versus structural phase transformations in compositionally complex alloys, npj Computational Materials, 7 (2021) 1-8.
[51] H. Chen, A. Kauffmann, S. Laube, I.-C. Choi, R. Schwaiger, Y. Huang, K. Lichtenberg, F. Müller, B. Gorr, H.-J. Christ, Contribution of lattice distortion to solid solution strengthening in a series of refractory high entropy alloys, Metallurgical and Materials Transactions A, 49 (2018) 772-781.
[52] Y. Zhang, Y.J. Zhou, J.P. Lin, G.L. Chen, P.K. Liaw, Solid-solution phase formation rules for multi-component alloys, Advanced engineering materials, 10 (2008) 534-538.
[53] M. Muzyk, D. Nguyen-Manh, K. Kurzydłowski, N. Baluc, S. Dudarev, Phase stability, point defects, and elastic properties of WV and W-Ta alloys, Physical Review B, 84 (2011) 104115.
[54] T.-K. Tsao, A.-C. Yeh, C.-M. Kuo, K. Kakehi, H. Murakami, J.-W. Yeh, S.-R. Jian, The high temperature tensile and creep behaviors of high entropy superalloy, Scientific reports, 7 (2017) 1-9.
[55] S. Shang, L. Hector Jr, Y. Wang, Z. Liu, Anomalous energy pathway of vacancy migration and self-diffusion in hcp Ti, Physical Review B, 83 (2011) 224104.
[56] Y.N. Osetsky, D. Bacon, N. De Diego, Anisotropy of point defect diffusion in alpha-zirconium, Metallurgical and Materials Transactions A, 33 (2002) 777-782.
[57] D. Nguyen-Manh, S. Dudarev, A. Horsfield, Systematic group-specific trends for point defects in bcc transition metals: an ab initio study, Journal of nuclear materials, 367 (2007) 257-262.
[58] A. Seeger, The study of point defects in metals in thermal equilibrium. I. The equilibrium concentration of point defects, (1973).
[59] D.A. Terentyev, L. Malerba, and M. Hou, Dimensionality of interstitial cluster motion in bcc-Fe, Phys. Rev. B 75 (2007) 104108.
[60] G.H. Vineyard, Frequency and isotope effects in solid state rate processes, J. Phys. Chetn. 3 (1957) 121-127.

# Supplemental Information

# Vacancy formation energies and migration barriers in multi-principal element alloys

Ankit Roy[a,b], Prashant Singh[b], Ganesh Balasubramanian[a], and Duane D. Johnson[b, c]

[a] Department of Mechanical Engineering and Mechanics, Lehigh University, Bethlehem, PA 18015, USA
[b] Ames Laboratory, United States Department of Energy, Iowa State University, Ames, IA 50011, USA
[c] Department of Materials Science & Engineering, Iowa State University, Ames, IA 50011, USA

**Additional information on measurement of vacancy formation energy**

The specific heat of a defect free crystal depends very weakly on temperature and witnesses negligible increase with rising temperature [40]. But in practicality, all metals especially the refractory metals undergo a non-linear increase in the specific heats with rising temperature [40]. The origin of this phenomenon could be cited to mainly two factors: anharmonicity and point-defect formation. A reliable method to separate the contributions from these two effects is by measuring the specific heats at high and low rates of temperature changes. At a high rate of temperature change, the defect concentration is not able to equilibrate to the temperature change and hence there is a linear increase in specific heat only due to anharmonicity, while at a low rate of temperature change the nonlinear increase in the specific heats is caused due to the formation of point-defects. Thus, the excessive rise in specific heat can be used to calculate the concentration of the point defects. In principle, calorimetric measurements can provide the vacancy formation energy by the dependence of vacancy contribution on temperature variation. However, ref. [58] concludes that a majority portion of the rise in specific heat with temperature is mainly due to the lattice anharmonicity and not due to vacancies hence specific heat analyses may not be an accurate method of studying vacancies.

**Atomic configurations for Ti vacancy calculations in stepped slab**: **(a)** the pristine Ti stepped slab with all atoms intact and coexisting with vacuum in the vacant volume. **(b) Case 1** an atom from the bulk moved to the hcp hollow position adjacent to the step edge mimicking migration of vacancy from bulk to surface due to absorption of energy. **(c) Case 2** an atom from the bulk moved to the hcp hollow position above the surface. The top view magnifies the position of the moved atom marked in green. **(d) Case 3** an atom from the bulk moved to the bridge position (center of the line joining the two atoms on the surface directly below, at a constant b vector in the diagram) above the surface. The top view magnifies the position of the moved atom marked in green. **(e) Case 4** an atom from the bulk moved to infinity replicating the case of highest energy requirement due to maximum number of bonds being broken. **(f) Case 5** an atom from the top surface extracted and moved to infinity replicating the case of lowest energy requirement due to minimum number of bonds being broken. **(g) Case 6** an atom at the step edge extracted and moved to infinity replicating the case of intermediate energy requirement due to less than the maximum number of bonds being broken.

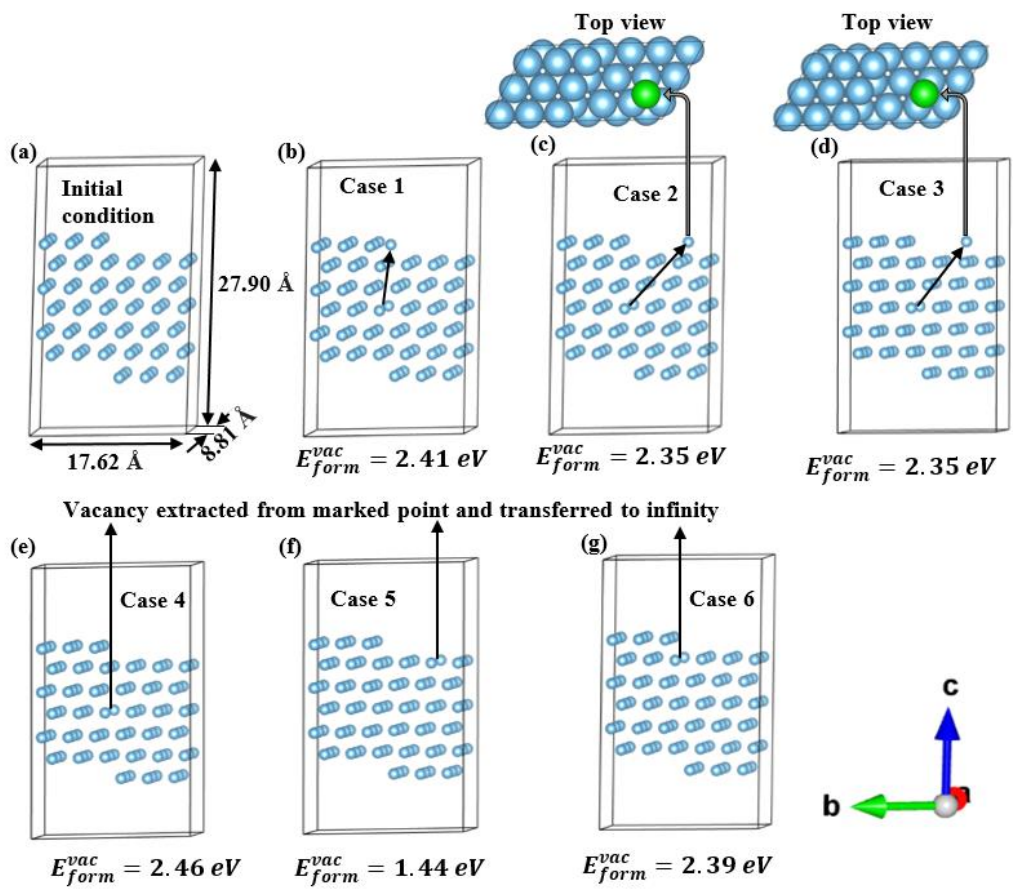

**Figure S1:** Atomic configurations detailing the surface energetics of 6 points of atom extraction visualized in a stepped slab structure of Ti.

**Effect of vacancy environment on local lattice distortion in MPEA-** The effect on neighboring atomic environment with(out) metal-vacancy defect quantified via vector atomic displacements for (a) Mo (Fig. S2), (b) Ta (Fig. S3), (c) W (Fig. S4) and (d) Zr (Fig. S5) atom/vacancy. The neighbors of Mo atom which are mostly Mo atoms, undergo small displacements as shown in (a) due to a purely metallic bonding existing between them. As shown in (b), Ta atom/vacancy in environment 2 shows a minor displacement of the surroundings when both first and second nearest neighbors comprise of Mo atoms. This is clearly due to the uniformity in the environment. The case reverses and higher displacements of neighboring atoms is noticed when there are unlike atoms in the surroundings. For the case of W (c), higher displacements are noticed when Ta and Zr are present as the first NNs, as seen in environment 1. This is due to the highly covalent nature of W-Ta bonds, as discussed in the main text. For the case of Zr (d), higher displacements are noticed in environment 1 where Ta, W and Ti are present as NNs. Because Ta and W are both $5d$ elements, their interaction with Zr ($4d$) causes a charge transfer to occur and results in a higher displacement in the NNs.

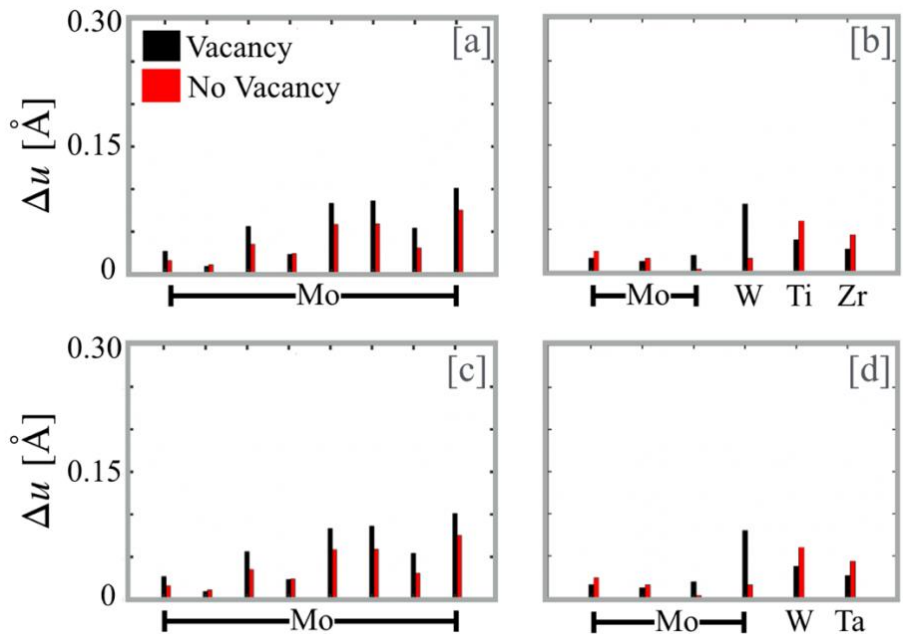

**Figure S2.** Vector atomic displacements in $(Mo_{0.95}W_{0.05})_{0.85}Ta_{0.10}(TiZr)_{0.05}$ of (a, c) first-near neighbor in environment 1, and (b, d) second-nearest neighbors in environment 2 around Mo vacancy (black bars) compared represent displacements for no-vacancy cases (red bars).

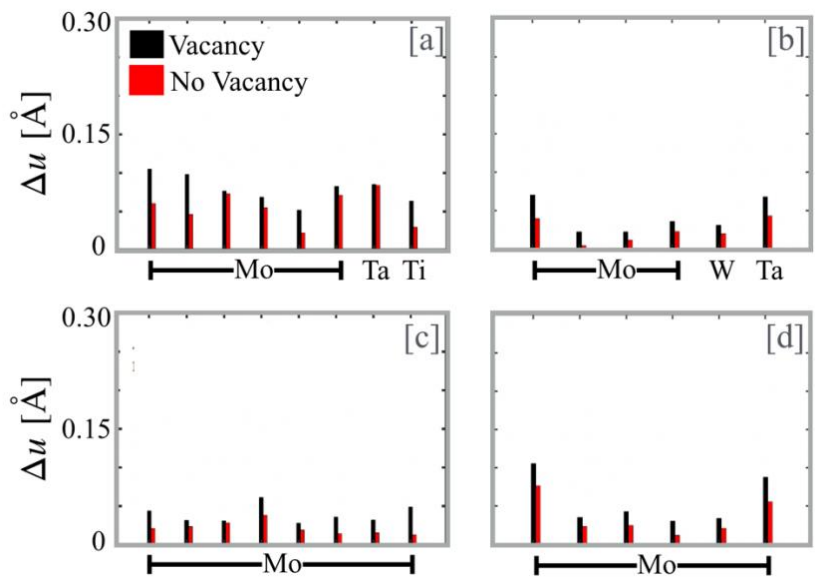

**Figure S3.** Vector atomic displacements in $(Mo_{0.95}W_{0.05})_{0.85}Ta_{0.10}(TiZr)_{0.05}$ of (a, c) first-nearest neighbor in environment 1, and (b, d) second-nearest neighbors in environment 2 around Ta vacancy (black bars) compared to no-vacancy case (red bars).

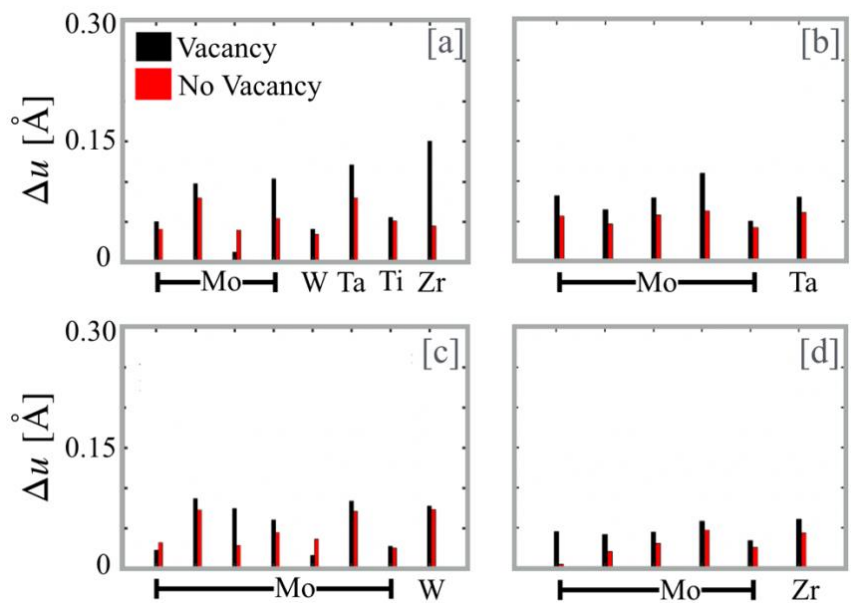

**Figure S4.** Vector atomic displacements of (a, c) first in environment#1, and (b, d) second nearest neighbors (NNs) in environment#2 around the W vacancy (black bars) compared to no-vacancy case (red bars) in $(Mo_{0.95}W_{0.05})_{0.85}Ta_{0.10}(TiZr)_{0.05}$.

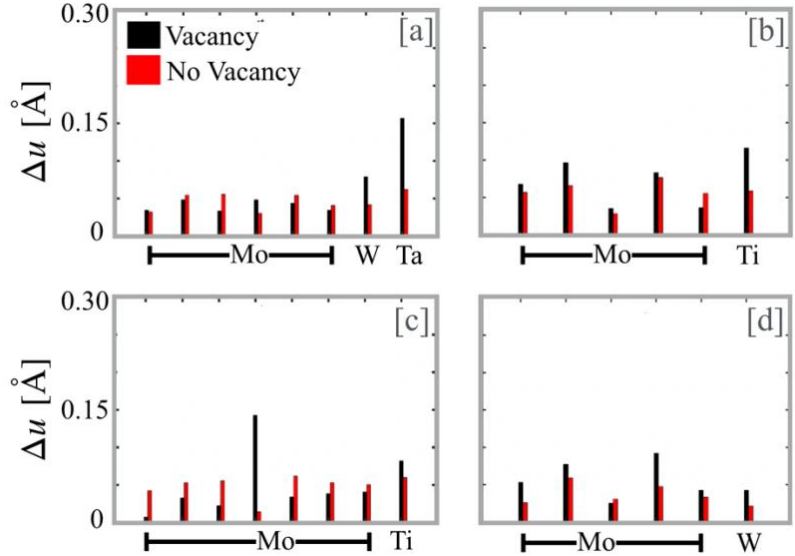

**Figure S5.** Vector atomic displacements in $(Mo_{0.95}W_{0.05})_{0.85}Ta_{0.10}(TiZr)_{0.05}$ of (a, c) first-near neighbor in environment 1, and (b, d) second-near neighbors in environment 2 around Zr vacancy (black bars) compared to no-vacancy case (red bars).

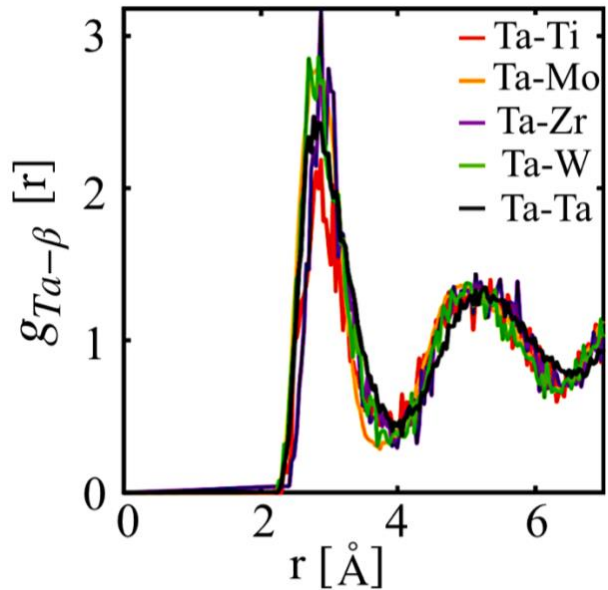

**Figure S6.** Pair-correlation functions $[g_{Ta\beta}(r)]$ for like and unlike atoms in $(Mo_{0.95}W_{0.05})_{0.85}Ta_{0.10}(TiZr)_{0.05}$. Unlike pairs (Ta-W) dominates over like pairs (Ta-Ta). This indicates strong affinity existing among unlike pairs that impedes motion of atoms in the MPEA lattice.